\documentclass[11pt]{article}
\usepackage[utf8]{inputenc}
\usepackage[T1]{fontenc}
\usepackage{geometry}
\usepackage{graphicx}
\usepackage{float}
\usepackage{caption}
\usepackage{hyperref}
\usepackage{xurl}
\usepackage{enumitem}
\hypersetup{colorlinks=true,linkcolor=black,urlcolor=blue,citecolor=black}
\newcommand{\tagline}[1]{\begin{center}\emph{#1}\end{center}}
\title{Short-Term Gain, Long-Term Fragility:\\AI Labor Substitution and the Erosion of Sustainable Capability}
\author{Dr. Wolfgang Rohde \\ AiSuNe Foundation Research \\ \texttt{inquiry@AiSuNe.com} \\ \url{https://aisune.com}}
\date{}
\begin{document}
\maketitle
\begin{abstract}
What looks like acceleration can be a quiet transfer of burden from the present to the future. Attempts to replace human labor with AI systems are often presented as rational responses to technological progress, but that view is incomplete and, in many cases, structurally short-sighted. Across software development and adjacent knowledge industries, corporate and political actors are increasingly attracted to AI because it appears to reduce labor costs, speed output, and improve short-term financial metrics. Yet these gains may be achieved by drawing down forms of human capability that are slow to build and difficult to restore. The paper develops a mechanism of capability masking and capability erosion under AI labor substitution. This dynamic can also be understood as a broader accumulation of technical, capability, and institutional debt. In this mechanism, AI-generated output creates a persuasive appearance that organizational capability has been replaced, even when the underlying dependence on skilled human labor remains. That appearance can support hiring restraint and deferred structural reform, while slower costs accumulate in the background. Empirical studies of AI-assisted coding show that generated output still requires substantial human verification and remains uneven in correctness, maintainability, and security [1-2]. Repository-level studies further suggest that current systems struggle to use broader repository context reliably in large codebases [3-6]. At the same time, evidence from labor-market research, political economy, and industrial-strategy analysis suggests that the substitution dynamic is being driven by managerial cost incentives and national competition, while creating additional risks of capital concentration and platform control over AI infrastructure and outputs [7-15]. The result is a system that appears more efficient in the short term while becoming more fragile over time. The deeper danger is not merely technical underperformance, but a broader pattern of capability erosion that could produce long-term economic fragility, concentration of power, and social disruption.

\textbf{Keywords:} AI labor substitution; software engineering; apprenticeship pipeline; capability erosion; institutional short-termism; political economy of AI.

\end{abstract}
\section{Introduction}
AI adoption is increasingly discussed in the language of inevitability. Firms face a steady mix of suggestion, benchmarking, and peer pressure to automate aggressively, reduce headcount where possible, and move quickly before competitors gain a strategic advantage. Political discourse in many countries often mirrors this tone. Governments celebrate innovation, promise productivity gains, and hesitate to challenge firms that frame labor substitution as modernization. In this climate, replacing humans with AI is often treated as a sign of strategic competence rather than as a decision with deep institutional and social consequences.

This framing is reinforced by visible moves among large firms. Microsoft cut about 3\% of its workforce in May 2025, or roughly 6,000 to 7,000 employees, while emphasizing organizational restructuring for a changing market; Amazon's CEO said in June 2025 that extensive use of generative AI and agents would reduce the company's total corporate workforce in the coming years; Salesforce said AI had reduced some of its hiring needs in engineering and customer service; and Anthropic, the company behind Claude Code, publicly reported very high internal AI code authorship, including about 90\% for Claude Code itself [16-19]. These examples do not prove a uniform one-to-one replacement of workers by AI. They do, however, show that AI is already being used as part of the managerial justification for workforce reduction, hiring restraint, and leaner staffing models.

The same logic is appearing in policy-facing corporate arguments. In testimony before the U.S. Senate Commerce Subcommittee on March 3, 2026, Siemens argued that industrial AI could help address a manufacturing labor shortage, citing more than 400,000 open jobs nationwide and stating that industrial AI "doesn't just solve the workforce shortage" but "changes the equation" [20-21]. Sector-specific shortages can coexist with broader labor-market weakness. But the juxtaposition matters: the Bureau of Labor Statistics reported a 4.4\% unemployment rate in February 2026, 7.6 million unemployed people, and a decline of 92,000 nonfarm payroll jobs in the same month [22]. The contradiction is therefore not that shortages cannot exist, but that AI is increasingly presented simultaneously as a remedy for scarcity and as part of a wider logic of labor reduction. In practice, similar narratives can be used to justify broader workforce reduction even when they originate in more specific sectoral constraints. That framing risks normalizing workforce displacement as a natural side effect of innovation and staying competitive even when its broader effect may be to narrow employment opportunities and compress human skill formation.

This paper argues that the dominant framing is too narrow. The issue is not simply whether AI tools can help people work faster. In many settings, they can. The more important question is what happens when organizations begin to treat AI not as a complement to human capability, but as a justification for replacing humans in the workforce and shrinking the human systems on which long-term capability depends. These human systems include apprenticeship and mentoring, tacit knowledge transfer, judgment formation, peer review, and the gradual development of domain expertise. They also include the social infrastructure that allows younger workers to enter professions, learn skills, build experience, and build stable lives. Here, capability means the durable human ability to interpret, maintain, adapt, and improve complex systems; institutional memory means the retained knowledge of how those systems evolved and why they work as they do; and tacit knowledge means the practical, often undocumented understanding acquired through participation rather than formal specification. The argument can also be read as a theory of accumulated debt. In software, technical debt refers to short-term expedients that defer complexity, cleanup, and repair into the future. Capability debt arises when visible output is preserved by weakening the human skill, review capacity, and apprenticeship structures needed to sustain reliable work over time. Institutional debt describes the deferred costs created when firms and states underinvest in governance and social systems on which long-run resilience depends. The discussion begins from software development, where the evidence is richest, and then extends by analogy to other apprenticeship-based domains in which skill is reproduced through supervised practice and institutional memory. Software engineering is emphasized here not because the mechanism is unique to that sector, but because AI effects on work are unusually observable and measurable in code repositories, tooling, and field studies, making it an early, well-documented case from which the mechanism can be examined by analogy in other apprenticeship-based industries. More generally, the underlying logic is not new. Scholarship on corporate short-termism has long linked short-horizon pressure to underinvestment in innovation and workforce training, even outside the AI context [23-24]. Management research on AI likewise warns that augmentation and automation are not cleanly separable categories, even though organizational rhetoric often prefers the language of augmentation [25]. The next sections clarify the paper's method and relation to adjacent literatures before turning to the mechanism itself.

\section{Approach}
\tagline{To understand a structural danger, one must follow the pattern across domains rather than isolate it within a single metric.}

This paper is a conceptual synthesis rather than a new empirical study. Its contribution is to identify and formalize a mechanism of capability masking and capability erosion by integrating several strands of recent evidence: empirical work on AI-assisted coding quality and productivity, repository-level and long-context studies, current labor-market reporting, policy testimony, and literature on AI-related incentives, labor displacement, and techno-solutionism.

The evidentiary strategy is selective and scoped. The paper relies primarily on peer-reviewed research, working papers and preprints, official testimony, government labor statistics, and major business reporting when discussing current corporate behavior or layoffs. It does not claim to measure the full magnitude of labor substitution directly, nor does it claim that software evidence alone is sufficient to establish a universal effect across sectors. Instead, it uses converging evidence to explain how short-term output gains, hiring restraint, verification burdens, and uncertainty about model maturity can interact as parts of a single mechanism that produces apparent efficiency while weakening long-term capability.

This method is appropriate because no single one of the paper's four source literatures sees the whole process. Software-engineering studies can show verification burdens and context failures, but not by themselves explain labor-market entry or platform concentration. Labor-market research can show substitution pressure and retraining limits, but not the concrete technical failure modes through which masking becomes plausible inside firms. Political-economy and geopolitical work clarify incentives, but not the day-to-day engineering conditions under which capability appears replaceable before it is actually reproducible. A conceptual synthesis is therefore necessary to identify the mechanism as a cross-domain pattern rather than as an isolated result within any single field. Methodologically, software is used here as a leading-indicator sector: AI adoption is early, effects are comparatively measurable, and changes in verification, workflow, and hiring practices become visible there sooner than in many other domains.

The paper also does not assume that the mechanism will unfold identically across jurisdictions. Labor protections, welfare institutions, industrial policy, and regulatory regimes can slow, redirect, or intensify different parts of the process. The claim is therefore not that every country will experience the same sequence at the same speed, but that the underlying mechanism is visible across settings wherever AI adoption, labor substitution pressure, and weak capability accounting interact.

\section{Related Work}
\tagline{No single literature sees the whole machinery; only their overlap reveals how technical, economic, and political forces lock together.}

This paper builds on four adjacent literatures. The first concerns AI-assisted software engineering, where recent work has examined code quality, maintainability, security, repository-level retrieval, and the productivity effects of coding assistants [1-5]. The second concerns AI and labor markets, including task exposure, changing labor demand, worker adaptability, and the limits of retraining under technological disruption [8-9,26-27]. The third concerns techno-solutionism and the political economy of AI, where scholars and policy analysts have examined how technical systems are presented as substitutes for slower forms of institutional reform [10-12]. The fourth concerns AI, industrial strategy, and geopolitical competition, where researchers and policy institutions examine AI as a strategic technology linked to national competitiveness, technological sovereignty, and rivalry over infrastructure and control [13-15].

The present paper does not attempt to replace these literatures with a new dataset or a single-sector causal estimate. Its contribution is instead integrative. It connects software-engineering evidence about verification, context limits, and non-surgical edits with labor-market evidence about substitution pressure and adaptation constraints, and then situates both within a broader mechanism of capability masking and capability erosion. It also treats geopolitical competition as an amplifier of that mechanism: if AI adoption is framed as a matter of national advantage, restraint becomes politically harder even when technical limitations and social costs remain substantial. Figure 2, introduced alongside the mechanism below, makes this relationship explicit. The point is not that the paper itself sits at the center, but that four adjacent literatures illuminate different parts of the same mechanism: software engineering clarifies technical limits, labor-market research clarifies substitution pressure, political economy explains the incentive structure, and geopolitics explains why caution is often politically costly. In that respect, the paper is closest to a conceptual model of organizational and social risk grounded in empirical findings from several neighboring domains.

\section{AI Labor Substitution as Institutional Short-Termism}
\tagline{When institutions mistake visible output for durable capability, speed becomes a solvent of memory.}

The argument developed here is that large-scale AI labor substitution reflects a familiar pattern of short-term extraction. More specifically, the paper describes a mechanism of capability masking and capability erosion. In this paper, masking refers to the perceptual and accounting-level phenomenon: visible AI output creates the appearance that capability has been replaced. Erosion refers to the slower structural consequence: the human systems that actually sustain capability are weakened over time. The mechanism can be stated in five linked steps. First, AI produces visible and plausible output that appears to substitute for human work. Second, managers and policymakers infer from that output that underlying capability has been replaced. Third, organizations may respond through hiring restraint, leaner staffing, and deeper dependence on external platforms. Fourth, hidden costs accumulate through verification burdens, loss of tacit knowledge, and contraction of the apprenticeship pipeline. Fifth, these firm-level decisions scale into broader fragility in the form of weaker organizational resilience, narrower entry paths, and more concentrated power. Looked at through the debt lens, the same sequence converts short-term convenience into a stack of deferred obligations: technical debt, capability debt, and institutional debt in the wider structures that reproduce skill and resilience. Figure 1 condenses this sequence into a single causal chain: rapid visible output sits at the front of the process, while the less visible losses in capability and resilience emerge later.

\begin{figure}[htbp]
\centering
\includegraphics[width=\textwidth]{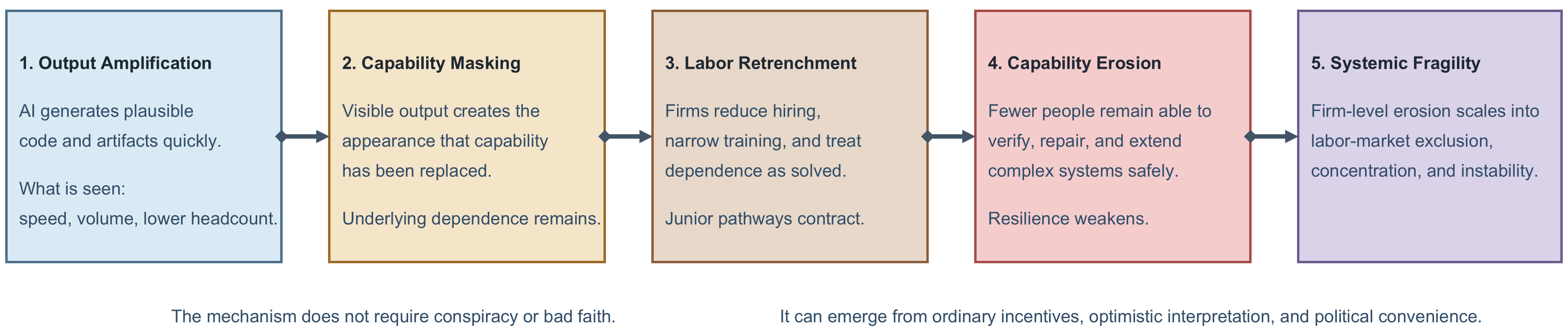}
\caption{Mechanism of capability masking and capability erosion under AI labor substitution.}
\label{fig:1}
\end{figure}

This is not a claim that AI is inherently harmful, nor that every form of automation is socially destructive. It is instead a claim about incentives, timing, governance, and power. The mechanism described here does not require bad faith or coordinated intent. It can emerge from ordinary incentive structures, optimistic interpretation of AI output, and the political and financial convenience of accepting visible productivity as proof of real substitution. There is also real evidence that AI can complement human work in bounded and well-governed settings, a point developed more fully later in the paper. The issue here is not whether beneficial use cases exist, but whether they justify substitution at a scale that weakens the human systems on which long-term capability depends.

Short-termism is a pattern in which institutions prioritize immediate, measurable gains over slower, more durable forms of value. In corporate settings, this often appears as cost cutting, labor reduction, and metric optimization pursued in ways that weaken future capability. AI intensifies this logic because it can be interpreted as evidence that a wide range of cognitive tasks no longer require the same depth of human participation.

This logic is attractive for several reasons. Labor is one of the most visible costs on a balance sheet, so replacing workers or freezing hiring produces rapid financial effects. Capital markets and competitive pressure also reward near-term gains more reliably than investments in apprenticeship, redundancy, or institutional memory.

The problem is that organizations do not operate on visible output alone. They rely on judgment, tacit coordination, informal review, and accumulated contextual knowledge. These are not easily captured in dashboards, but they are central to resilience. When firms reduce human participation because AI appears to reproduce narrow outputs, they risk confusing the surface substitution of artifacts with capability substitution in judgment and understanding. That is the masking stage of the mechanism: visible performance is overread as evidence that the underlying capability problem has been solved. The erosion stage follows later, as the human foundations of actual competence are allowed to weaken.

Repository-level studies likewise show that realistic multi-file development depends on retrieving and using cross-file context rather than only local file state, and that this remains a nontrivial evaluation and engineering problem [3-4]. More generally, long-context research shows that model performance degrades when relevant information must be recovered from less salient parts of the context window rather than from obvious positions near the beginning or end [28]. These technical limitations make it risky to assume that current systems can substitute for developers' ability to navigate non-salient but crucial context over the life of a codebase.

\vspace{0.5\baselineskip}
\begin{figure}[htbp]
\centering
\includegraphics[width=\textwidth]{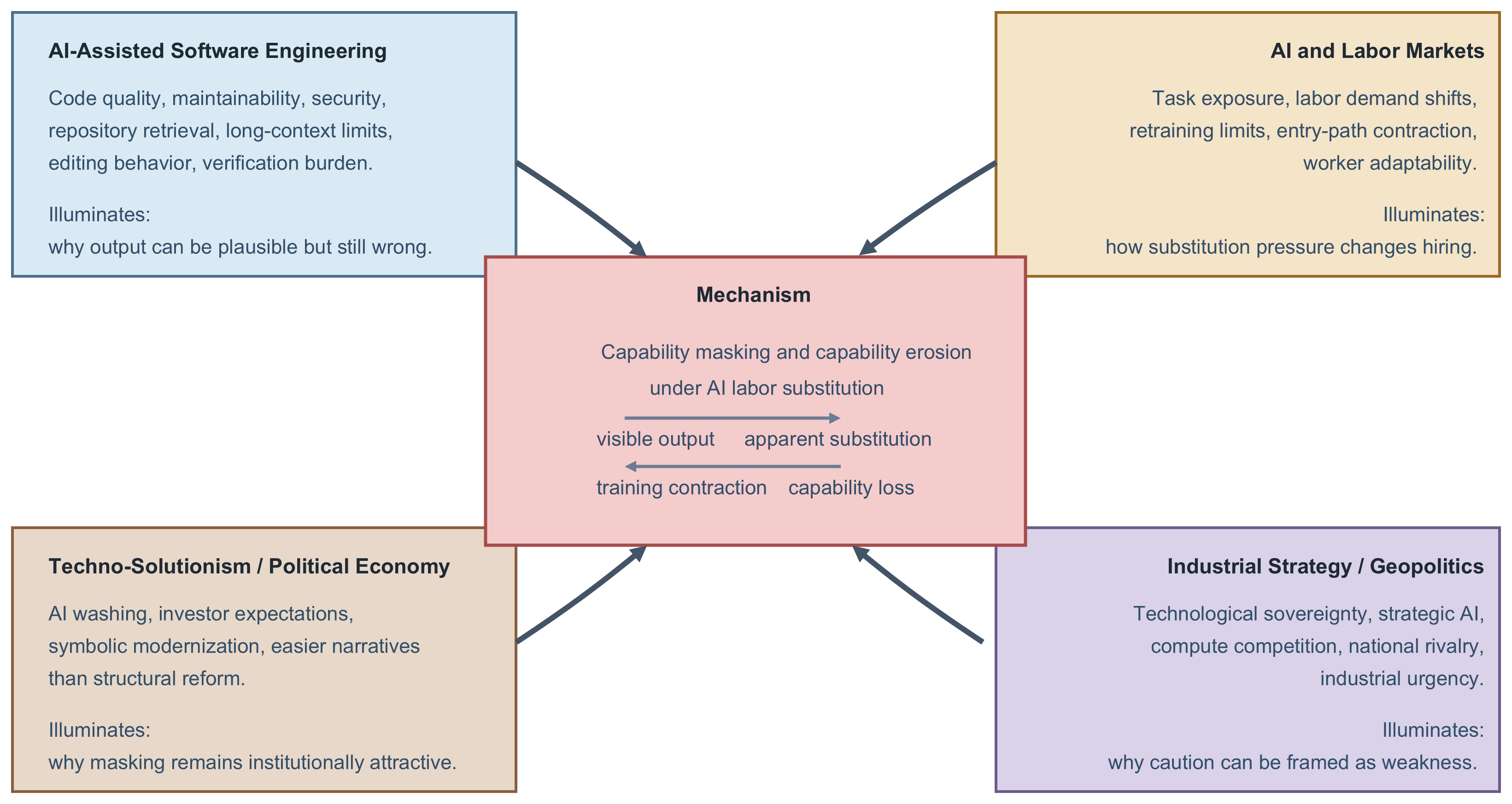}
\caption{Related work map: four literatures illuminating different dimensions of the mechanism of capability masking and capability erosion.}
\label{fig:2}
\end{figure}

In software engineering, this distinction is especially important. Research on AI-assisted code generation consistently shows that generated code remains uneven in correctness, maintainability, and security, and that human verification remains essential [1-2]. Security-focused work further shows that generated code can introduce significant vulnerabilities even when it appears productive in the short run [2].

These findings do not imply that AI has no value. They imply that AI often shifts effort rather than eliminating it, and that removing humans on the assumption of full substitution may weaken the very oversight that keeps complex systems dependable.

\section{The Hidden Loss of Organizational Capability}
\tagline{Organizations often discover the value of human judgment only after they have made it scarce.}

The deeper danger in replacing humans with AI is the erosion of organizational capability: more than task completion, it includes the ability to diagnose novel failures, interpret ambiguous signals, understand system history, adapt under pressure, and train the next generation of practitioners. These capabilities emerge through sustained human participation, not merely through the presence of functional artifacts. In the debt language introduced above, this is where capability debt becomes visible: immediate output is maintained by borrowing against future human understanding.

The transition from masking to erosion matters here. Masking is the earlier interpretive error: plausible output is mistaken for durable capability. Erosion is what follows when institutions act on that mistake and allow the human systems underneath real capability to thin out.

Organizations often discover the value of human capability only when something breaks. In routine conditions, AI systems may appear to perform adequately, especially when tasks are repetitive or strongly structured. Under stress, however, resilience depends on people who understand why systems behave as they do, how historical decisions shaped current architectures, and where hidden failure points are likely to emerge. Firms that hollow out their human base may preserve throughput while losing recovery capacity.

This matters because modern firms depend on layered and opaque technical systems. The cost of losing experienced people is therefore not only the loss of current productivity but the loss of interpretive depth. AI tools can generate plausible answers and plausible code, but plausibility is not the same as understanding. If organizations replace humans faster than they develop methods for preserving judgment and institutional memory, they may create a brittle operating model in which fewer people truly understand the systems on which the business depends.

Recent field evidence makes this problem concrete. METR's 2025 randomized controlled trial of experienced open-source developers working on their own large repositories found that developers were 19\% slower with AI tools, even though they believed they had been sped up. Developers reported that AI performed worse in large and complex environments, accepted fewer than 44\% of generations, often had to make major changes to clean up AI code, and that models did not utilize important tacit repository knowledge or context [5]. Participants described the model as missing backward-compatibility requirements and introducing as many errors as it fixed [5]. This should not be read as a universal effect size for all developers or firms; it is evidence from experienced developers in a specific large-repository setting. Its value for the present argument is not that it settles the productivity question everywhere, but that it shows how quickly visible gains can become misleading in environments where tacit context matters.

These observations closely match a broader engineering complaint about AI coding in large projects. The failure mode is often not that the model produces obviously nonsensical output. It is that it produces locally plausible output while optimizing against an incomplete representation of the system. This can lead to globally counterproductive decisions, especially where behavior depends on undocumented edge cases, compatibility guarantees, or architecture-level conventions. Figure 3 illustrates that point more precisely: a change can satisfy the immediate prompt and even local tests while still violating non-local constraints such as backward compatibility, auditability, or performance budgets. It can also lead to non-surgical editing behavior. OpenAI's description of Canvas is used here illustratively rather than as primary evidence: the product write-up notes that editing behavior required explicit tuning to balance targeted edits against whole-document rewrites, and that users may need to select specific sections to focus the model's attention [6]. Figure 4 complements this by contrasting a surgical change with a hammer-style rewrite that expands beyond the actual problem boundary. As one illustrative case suggests, broad rewrites versus focused modifications remain an active control problem rather than a solved property of model behavior.

\begin{figure}[htbp]
\centering
\includegraphics[width=\textwidth]{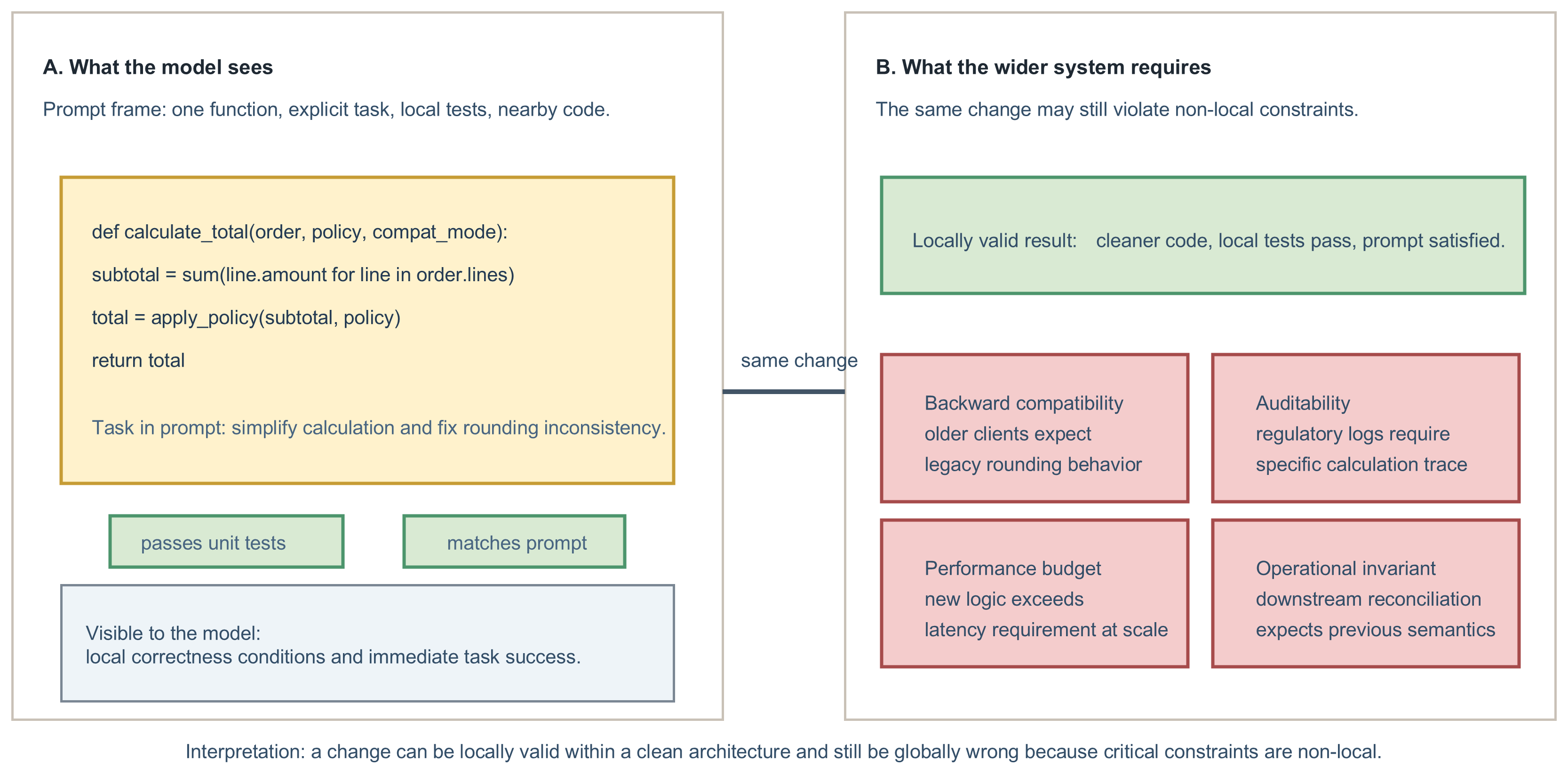}
\caption{A locally valid change can violate non-local constraints in a large system.}
\label{fig:3}
\end{figure}

This pattern is visible in current discussions of AI-assisted software development. Surveys and industry studies report higher throughput, but they also report trust gaps, verification bottlenecks, and instability when governance does not keep pace [29-30]. DORA in particular frames AI as an amplifier of existing practices rather than as a uniformly positive force: organizations with strong engineering discipline may benefit, while weaker organizations may simply accelerate existing dysfunctions [29]. This matters for labor substitution because removing humans in a poorly governed environment does not create efficiency. It reduces the stock of people capable of correcting amplified mistakes.

Where similar apprenticeship-based training structures exist, the same dynamic extends beyond software. The weakening of verification capacity, tacit knowledge transfer, and supervised learning is not unique to code; it is a broader organizational problem wherever skill is reproduced through guided practice rather than instant replacement.

\vspace{0.5\baselineskip}
\begin{figure}[htbp]
\centering
\includegraphics[width=\textwidth]{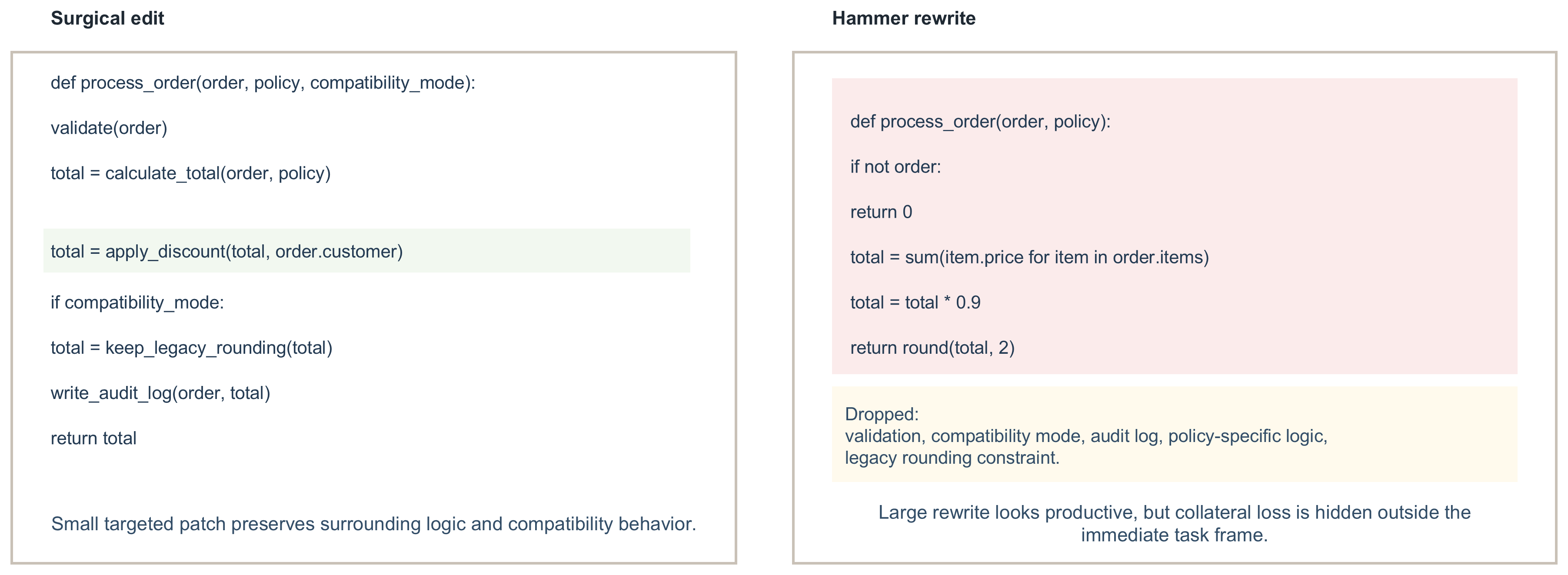}
\caption{Surgical edit versus hammer rewrite in AI-assisted coding.}
\label{fig:4}
\end{figure}

\section{Counterevidence and Boundary Conditions}
\tagline{Complementarity is real, but it is not the same thing as safe substitution.}

The strongest counterargument is that AI often complements rather than replaces human capability. That claim has real support. DORA reports that many teams using AI-assisted development report higher throughput and positive productivity effects on average, especially where engineering practices are already strong [29]. GitHub has likewise reported modest but statistically significant quality improvements in controlled Copilot-assisted tasks, including gains in readability, reliability, maintainability, and test pass rates [31]. Because that study was vendor-produced, it should not be treated as neutral proof on its own, but it remains relevant as evidence that complementarity claims are not purely rhetorical. These findings matter because the paper's argument is not that AI has no value. It is that value at the level of assisted task completion does not by itself justify workforce replacement or the contraction of training capacity.

The boundary condition is therefore crucial. Positive results are most convincing in settings where AI is used as a complement inside already disciplined environments, where human review remains intact, and where the task is sufficiently bounded for local success to map reasonably well onto broader system goals. The present argument is that once organizations move from complementarity to substitution, the risk profile changes. The same tools that can help strong teams move faster can become harmful when they are used to justify thinner staffing, weaker apprenticeship, and overconfidence about what visible output really proves. In that sense, the paper does not deny beneficial use cases; it argues that those cases do not settle the question of sustainable labor replacement.

\section{The Collapse of the Apprenticeship Pipeline}
\tagline{Every generation of expertise depends on a previous generation willing to remain slower for a while.}

One of the most serious long-term risks of AI-driven labor substitution is the weakening of the apprenticeship pipeline. Complex professions reproduce themselves through structured exposure: junior workers begin with constrained tasks, learn from review and mentorship, and gradually acquire the judgment needed for senior responsibility. If organizations eliminate or sharply reduce junior roles because AI can perform parts of beginner-level work, they may reduce short-term costs at the expense of the future supply of experienced practitioners.

In knowledge-intensive sectors, the junior layer is the training ground from which future maintainers, architects, managers, and technical leaders emerge. A firm that cuts junior hiring for several years is not simply reducing costs in the present; it is reducing the number of people who will possess internal knowledge and mature judgment in the future. At industry scale, the same pattern can create a generational thinning of expertise. Figure 5 makes the temporal structure visible: the damage does not appear immediately in output metrics, but later in the diminished supply of experienced practitioners.

Recent reporting suggests that this risk is no longer hypothetical. Multiple sources have noted weakening demand for junior developers and growing concern that AI-assisted workflows are being used to justify reductions in entry-level hiring [7]. More broadly, labor-market research finds that tasks with higher AI exposure subsequently experience reduced labor demand, and that retraining pathways for AI-exposed workers are uneven rather than automatic [8-9]. These findings do not isolate junior software roles on their own, but they align with a structurally plausible mechanism: when firms can automate entry-level tasks, they may preserve senior labor while shrinking the intake through which future senior labor is formed.

Anthropic's own public messaging illustrates the tension clearly. Fortune reported in January 2026 that Boris Cherny, head of Claude Code, said 100\% of his own code was written by Claude and that an Anthropic spokesperson put company-wide AI-generated code at 70\% to 90\%, with about 90\% for Claude Code itself [19]. Because Anthropic is both the developer of Claude Code and the source of the claim, these figures should be read with appropriate caution. The point is not that they provide neutral proof of substitution, but that they show how quickly organizations can normalize very high AI authorship rates. Even in the same reporting, however, the remaining human role is still essential: prompting, selecting, reviewing, editing, and integrating outputs into real systems. The result is not the disappearance of human dependence, but its migration upward toward more experienced oversight.

History offers a useful caution. Legacy-system modernization has repeatedly shown how expertise shortages become acute when critical systems depend on aging technologies and too few people possess the knowledge needed to maintain them. Recent GAO reporting notes that federal agencies continue to operate mission-critical legacy systems using languages such as COBOL and assembly, and that shortages of staff with specialized skills create significant mission risk and increase modernization costs [32]. The point of the analogy is not technological similarity. It is institutional memory. Systems that are neglected, poorly documented, or maintained by shrinking expert communities eventually reveal the cost of underinvestment in people. If AI-era firms erode their training pipeline while increasing dependence on complex digital infrastructure, they may reproduce a similar pattern on a broader scale, this time with AI-mediated systems whose behavior is even less transparent to non-experts.

\begin{figure}[htbp]
\centering
\includegraphics[width=\textwidth]{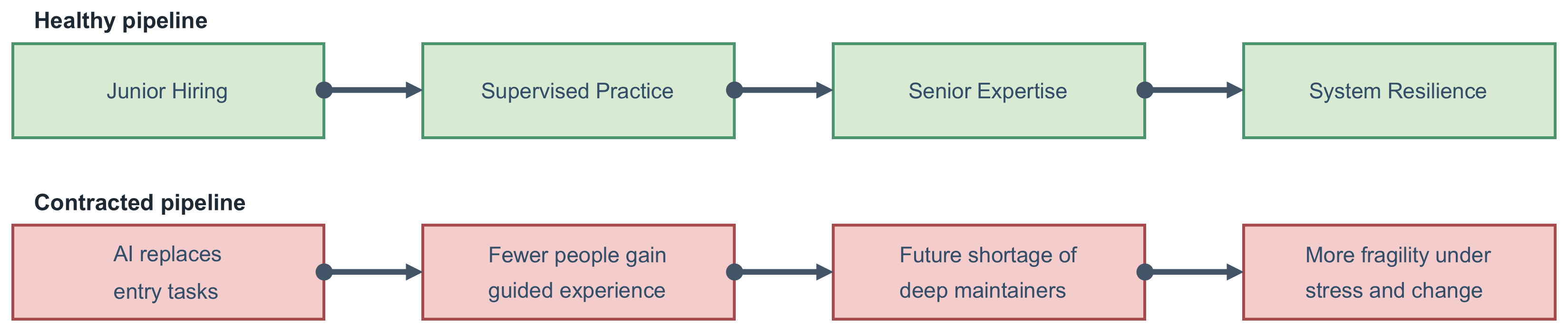}
\caption{Apprenticeship pipeline contraction and delayed expertise loss.}
\label{fig:5}
\end{figure}

This also helps explain why older automation analogies are incomplete. Earlier waves such as ATMs, office software, or industrial automation often displaced specific tasks while still leaving wider entry paths into the surrounding occupation. The point is not that there is a fixed amount of work to be done and that any automation must therefore destroy jobs. It is that the present wave is different in one important respect: it is frequently aimed at cognitive and entry-level work that functions as the training layer through which future expertise is formed. That does not mean reabsorption is impossible, but it does mean it cannot be assumed. Where automation narrows the apprenticeship channel itself, the long-run effect may be less substitution of one task for another than erosion of the occupational pipeline that once regenerated skill.

The consequences of AI labor substitution are not confined to firms. When enough organizations pursue the same short-term strategy, private decisions accumulate into public risk. A labor market with fewer entry-level openings does not merely inconvenience recent graduates. It narrows the path into stable, skilled work, delays household formation, increases economic insecurity, and weakens confidence in the social contract. Societies depend on the belief that effort, learning, and participation can lead to advancement. If that pathway is blocked at scale, the effects are economic, cultural, and political.

The social danger is therefore broader than headline unemployment. It includes underemployment, stalled skill formation, and the creation of a cohort that is formally educated but excluded from high-value sectors. This is especially destabilizing because digital industries have been framed for decades as engines of upward mobility. If AI is used to close entry ramps into those sectors while concentrating gains among capital owners, platform providers, and a small layer of already-established experts, inequality may widen in ways that are both material and symbolic.

\section{From Firm-Level Fragility to Social Risk}
\tagline{What many firms call efficiency in isolation can become fragility when repeated across a society.}

There is also a governance dimension. Political systems frequently respond slowly to structural labor shifts, especially when those shifts are presented as innovation successes. By the time displacement becomes visible in tax receipts, social service demand, household instability, or political anger, the institutional damage may already be deep. The danger is therefore a foreseeable systemic fragility ignored because incentives reward extracting immediate gains over cultivating long-term resilience.

This dynamic also raises questions of power concentration. AI capability is expensive to develop, compute-intensive to run, and increasingly mediated by a relatively small number of firms. If human labor is displaced while productive power becomes concentrated in a narrow platform layer, then society risks becoming more economically dependent on actors whose incentives are not aligned with broad social sustainability. The problem is therefore not only job loss, but the reorganization of power around infrastructures that are privately governed and unevenly accountable. Those infrastructures can also shape mediated knowledge itself: whoever controls widely used models and their distribution channels may influence not only production capacity but the examples, answers, and framings through which users encounter the world.

There is a further reason to question the confidence with which full replacement narratives are presented: it remains unclear whether current AI development paths will mature enough to deliver the level of reliability, contextual understanding, and consistency that those narratives assume. This technical uncertainty matters directly to the labor-substitution story, because firms often behave as if substitution quality will improve smoothly and predictably over time. That assumption underwrites the second step of the mechanism, in which visible output is overread as proof that capability has already been replaced. One source of uncertainty is the data pipeline itself. Recent work by Epoch AI estimates that the effective stock of high-quality, human-generated public text suitable for training frontier models is finite and may be largely utilized between 2026 and 2032 under current scaling trends, or earlier under data-intensive training regimes, but the authors present these as scenario-dependent projections rather than fixed forecasts [33]. At the same time, research in \emph{Nature} shows that indiscriminate recursive training on model-generated data can produce "model collapse," in which systems progressively lose information about the true underlying distribution and converge toward degraded outputs [34]. More recent work on synthetic data suggests that carefully mixed synthetic data can sometimes help in specific settings, but also confirms that pure or poorly designed synthetic training regimes can degrade downstream performance and require substantial caution [35]. Figures 6 and 7 summarize the broader implication. Figure 6 shows the core asymmetry of the paper: visible gains in speed or cost can rise early while capability and resilience erode later. Figure 7 makes the same point in layered form by distinguishing the visible surface of output from the hidden dependence on human review, training, and institutional competence. The implication is not that AI progress must end, but that the path to the maturity required for large-scale labor substitution is substantially more uncertain than current business rhetoric often suggests.

Once that broader fragility is visible, the next question is why institutions continue choosing the short-term path even when the longer-run costs are becoming easier to see.

\begin{figure}[htbp]
\centering
\includegraphics[width=\textwidth]{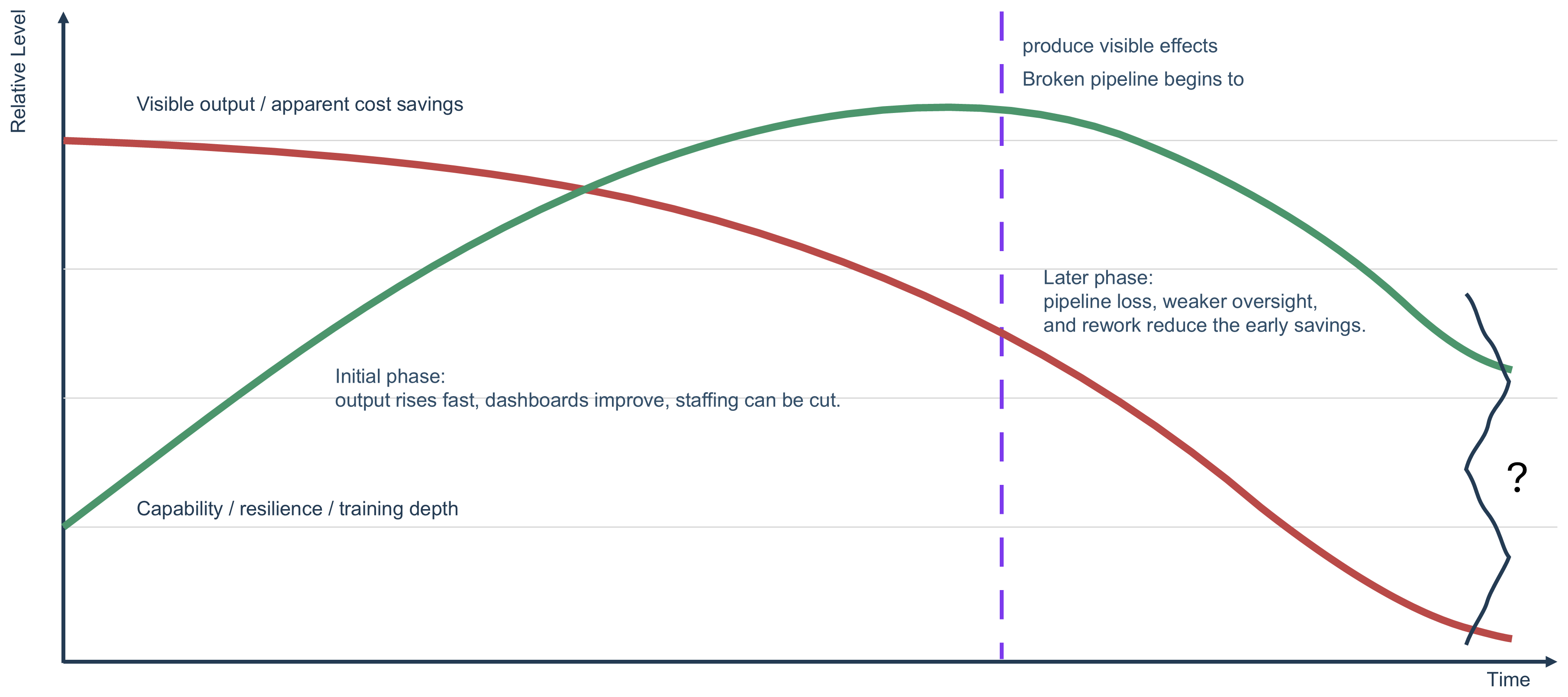}
\caption{Short-term visible gain versus long-term capability and resilience loss.}
\label{fig:6}
\end{figure}

\section{Why Management and Politics Keep Choosing the Short Term}
\tagline{Systems drift toward short-termism not only because it is profitable, but because it is easier to narrate than restraint.}

Understanding why this pattern persists requires moving from its systemic effects back to the incentives of the actors who reproduce it. The mechanism depends on decision-makers repeatedly choosing short-term gains over capability preservation. Part of the answer lies in incentive design. Managers are often rewarded for near-term cost reductions, visible productivity gains, and narratives of innovation. Politicians are rewarded for associating themselves with technological progress, investment flows, and national competitiveness. Neither group is consistently rewarded for preserving apprenticeship systems, maintaining labor-market entry pathways, or defending institutional redundancy.

Another reason is epistemic asymmetry. The gains from AI substitution are immediate, countable, and easy to publicize. The losses are delayed, diffuse, and harder to attribute. A hiring freeze can be measured this quarter; the absence of experienced maintainers five years later cannot be traced so cleanly to a single decision, and the slow erosion of tacit knowledge remains abstract until a crisis makes it concrete.

\vspace{0.5\baselineskip}
\begin{figure}[htbp]
\centering
\includegraphics[width=\textwidth]{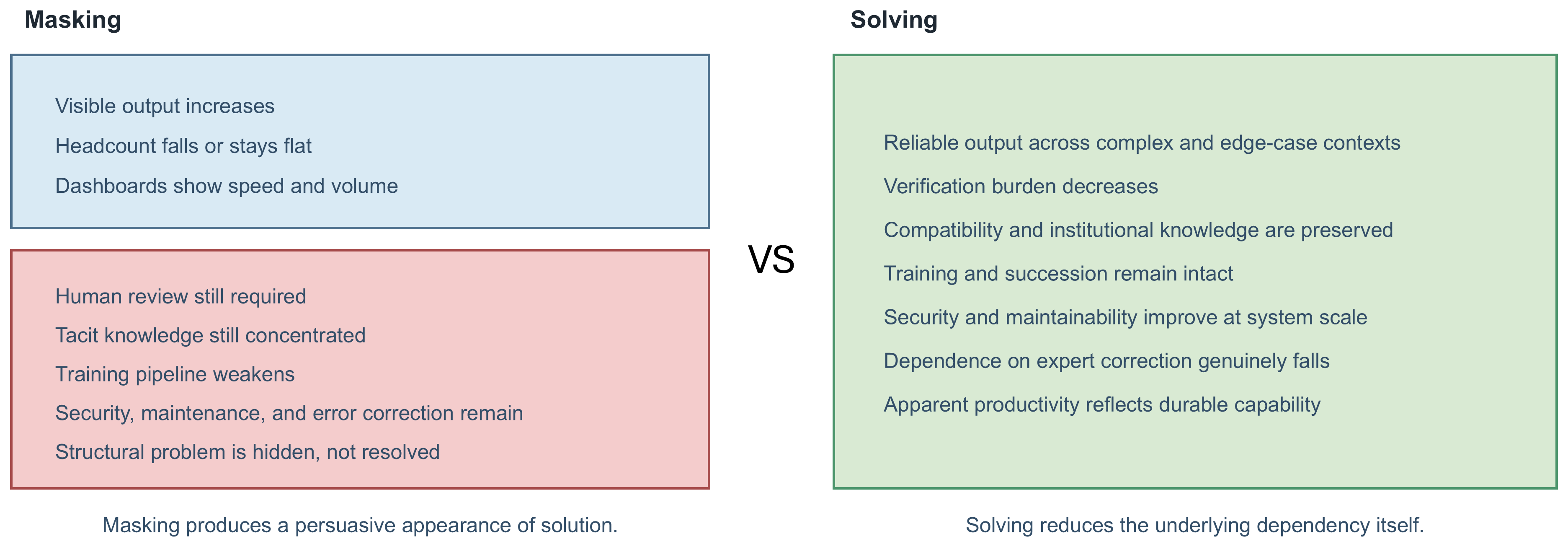}
\caption{Masking versus solving: visible output above unresolved structural dependence.}
\label{fig:7}
\end{figure}

There is also a political-economic component. Large firms have strong incentives to present AI as a substitute for labor because the promise of automation supports valuation, investor confidence, and cost-reduction narratives. Survey work by PwC and others shows strong expectations that AI will increase productivity, revenue, and profitability, while financial reporting has repeatedly warned that AI expectations may run ahead of actual capability and delivery [12]. At the same time, regulators have begun acting against "AI washing," underscoring that firms have incentives to overstate or strategically market AI capabilities in order to capture financial advantage [11].

Power concentration also matters at the platform layer itself. There is no single uncontested count of "true" LLM platform providers, because some firms supply closed model APIs, some release open-weight models, and others mainly distribute models through cloud ecosystems. Still, recent mappings make the concentration clear. The 2025 Foundation Model Transparency Index scored 13 major foundation model companies, while the most entrenched policy and market cluster it highlighted remained narrower still, including firms such as Amazon, Anthropic, Google, Meta, and OpenAI [36]. OECD analysis likewise warns that foundation model and AI infrastructure markets are prone to concentration through control of compute, cloud distribution, data, and complementary services [13]. The narrowness of the stack is also visible in formal partnerships and capital ties: Microsoft stated in January 2025 that core elements of its OpenAI relationship remained in place through 2030, including revenue sharing, OpenAI API exclusivity on Azure, continued investment, and a right of first refusal on new capacity [39]. Reuters later reported that Nvidia would invest in OpenAI while also supplying the data-center chips on which OpenAI's expansion depends, making the overlap between model providers, cloud partners, and compute suppliers even more explicit [40]. These examples do not exhaust the market, but they make the concentration pattern concrete. As with earlier cloud-platform choices, dependence on a generative-AI provider can also create forms of vendor lock-in through APIs, workflow integration, fine-tuning pipelines, surrounding tooling, and accumulated organizational dependence on one provider's interface layer. In some respects that lock-in may be deeper than ordinary cloud dependence, because replacing or retraining the model layer is often more expensive, technically demanding, and organizationally disruptive than relocating application workloads across infrastructure providers. If current trends continue, labor displacement may coincide with a deeper monopolization of technical power and operational control.

This concentration has an epistemic dimension as well as an economic one. Whoever controls widely used models, their system prompts, retrieval layers, and deployment channels can influence what kinds of answers are surfaced, which examples are privileged, and which topics are constrained or redirected. Concerns about DeepSeek led Taiwan to ban government use of the system, citing both censorship and information-security risks [37]. xAI likewise acknowledged in May 2025 that an unauthorized prompt modification caused Grok to push a specific political topic in unrelated conversations, illustrating how governance choices at a single provider can shape the narrative delivered by a model at scale [38]. The broader point is that control over widely used models can become partial control over mediated narrative and example space.

Political actors face a parallel temptation. Symbolic alignment with AI can serve as a visible performance of modernization, competitiveness, and national ambition, while the slower work of education reform, workforce development, labor-market adjustment, and institutional capacity building receives less attention. In this sense, AI can become attractive not only because it may solve problems, but because it offers a way to avoid confronting them directly. Brookings has argued that AI-driven labor displacement exposes the limits of simple retraining narratives, while recent scholarship on technological solutionism provides a conceptual language for understanding why technical systems are repeatedly presented as substitutes for deeper structural reform [10,27].

Geopolitical rivalry intensifies this temptation. When AI is framed as a strategic technology tied to industrial policy, technological sovereignty, and national power, slowing down to address capability loss can be portrayed as geopolitical weakness rather than institutional prudence [13-15]. These are related but distinct pressures: short-termism is temporal, techno-solutionism is epistemic and political, and geopolitics is strategic. In the mechanism itself, geopolitics operates less as a separate step and more as a modifying condition: it increases the pressure behind the second and third stages by making rapid adoption, external dependence, and labor reduction easier to justify. These pressures can also be stabilized by ecosystem structure: capital, compute demand, platform dependence, and infrastructure revenues can circulate inside a concentrated provider stack, reinforcing the appearance of momentum and making restraint even less attractive. This remains a loop among categories occupied by only a small number of dominant providers rather than a broad and competitive market.

Many institutions also remain attached to a model of progress in which replacing labor with technology is assumed to be inherently modernizing. This view treats human labor primarily as cost rather than as a source of adaptive intelligence, social stability, and institutional continuity. Under that assumption, the erosion of human systems can be interpreted as efficiency even when it is better understood as liquidation.

In software and other knowledge fields, the rhetoric of inevitability is particularly powerful. If AI appears capable of performing tasks once associated with trained professionals, then resistance to labor substitution can be framed as sentimental, anti-innovation, or economically naive. Yet such framing obscures a basic point: sustainable institutions are not built only by minimizing labor inputs. They are built by preserving the human capacities that allow systems to be understood, repaired, governed, and improved over time.

\section{Toward a Sustainable Alternative}
\tagline{The real test of progress is not how quickly labor can be removed, but whether capability can deepen without becoming narrower.}

A sustainable AI strategy would begin from complementarity rather than substitution. The relevant question is not how quickly humans can be removed, but how AI can extend human capability without destroying the training and institutional structures on which future capability depends. In software development, that means treating AI as an assistive layer within strong review, testing, and training processes rather than as a justification for shrinking the human core.

At the firm level, this implies preserving the apprenticeship pipeline, investing in mentorship, and measuring AI adoption partly by its effect on learning, resilience, and maintainability rather than only by cost savings or output speed. Organizations should ask whether AI use is increasing or decreasing the number of people who genuinely understand their systems. If the answer is that understanding is becoming narrower even while output rises, the firm may be becoming more fragile rather than more capable. More concretely, firms can maintain or increase junior hiring in AI-intensive teams, dedicate verification time to AI-generated output, and require that AI-assisted work be incorporated into post-mortem learning rather than treated as disposable throughput.

At the policy level, sustainability requires more than innovation rhetoric. It requires attention to labor-market entry, retraining capacity, market concentration, and the public consequences of widespread capability erosion. Policymakers should be skeptical of claims that aggregate productivity gains automatically translate into broad social benefit, especially when the gains are achieved through reduced employment opportunity and increased dependence on concentrated private infrastructure. It also requires scrutiny of who controls the models, compute, and distribution channels through which AI-mediated knowledge is produced, because technical dependence can become dependence on privately governed narrative infrastructure. Illustrative policy levers include reporting requirements when AI-related workforce reductions cross predefined material thresholds, disclosure of AI-linked headcount changes and training expenditures, procurement preferences for vendors that maintain apprenticeship capacity, and competition rules aimed at limiting excessive concentration in AI platforms and infrastructure. One practical model is the broader family of material-disclosure frameworks already used in corporate and public reporting: institutions are often required to disclose developments that materially affect risk, operations, or investor understanding. Similar logic could be adapted to large AI-linked workforce reductions, significant cuts to training capacity, or major increases in operational dependence on external AI platforms. The challenge is not inventing governance from nothing, but determining which changes are materially significant, how thresholds should be defined, and how reporting can be made resistant to evasion.

None of these levers is frictionless. Threshold design can be gamed, procurement rules can be captured, and concentration remedies are harder to enforce when supply chains and model access are globally distributed. But implementation difficulty is not an argument for passivity; it is part of the reason governance must be treated as a substantive design problem rather than as an afterthought.

Worker agency also matters. Sustainability does not depend only on better decisions by managers and states; it also depends on whether workers, professional communities, and collective institutions retain enough voice to shape how AI is introduced. Collective bargaining over deployment terms, consultation rights around workflow redesign, and professional norms that require meaningful human review can all function as counterweights to substitution pressure. If workers are treated only as objects of replacement rather than participants in governance, one more stabilizing layer of the system is removed.

The broader principle is simple: not every cost reduction is a social improvement, and not every efficiency gain is sustainable. A society that treats human capability as disposable may discover too late that it has optimized away the very capacities needed for resilience, legitimacy, and long-run prosperity. The practical goal is therefore not only to avoid technical debt in code or infrastructure, but also to avoid capability debt in organizations and institutional debt in the systems that train, govern, and stabilize complex work.

\section{Conclusion}
\tagline{The deepest cost of false efficiency is paid when a society no longer knows how to sustain what it still depends on.}

Short-term gains from AI labor substitution are frequently purchased by drawing down the very human capabilities that make complex systems governable. This paper has described that process as a mechanism of capability masking and capability erosion under AI labor substitution: organizations and political systems capture immediate savings and visible gains while weakening the human foundations of long-term capability. In that sense, the paper is also describing a broader accumulation of debt: technical debt in artifacts and systems, capability debt in the human layer that maintains them, and institutional debt in the wider structures that make resilience possible.

In software development, the evidence already shows that AI-generated output remains uneven and still depends heavily on human judgment, verification, and maintenance. More broadly, the reduction of junior pathways threatens the reproduction of expertise, while large-scale labor substitution risks producing wider social consequences in the form of exclusion, concentration, and instability. The central danger is therefore not simply that some firms may overestimate the abilities of AI tools. It is that entire institutions may begin to consume their future capacity in exchange for present convenience.

If that pattern continues, the eventual shock will not appear as a single technical failure. It will appear as a society that is less capable of training people, less capable of maintaining complex systems, less capable of distributing economic participation, and less capable of governing the infrastructures on which it depends. It may also appear as a society in which productive capacity, informational mediation, and narrative power are concentrated in too few private hands. The final question is whether societies can remain sustainable while treating human capability itself as expendable.

\section*{References}
\begingroup
\small
\sloppy
\begin{enumerate}[label={[\arabic*]}, leftmargin=0.7in, itemsep=0.6em]
\item Yetiştiren, B., et al. (2023). \emph{Evaluating the code quality of AI-assisted code generation tools}. arXiv. \url{https://arxiv.org/abs/2304.10778}
\item Fu, Y., et al. (2023). \emph{Security weaknesses of Copilot-generated code in GitHub projects: An empirical study}. arXiv. \url{https://arxiv.org/abs/2310.02059}
\item RepoCoder. (2023). \emph{RepoCoder: Repository-level code completion through iterative retrieval and generation}. arXiv. \url{https://arxiv.org/abs/2303.12570}
\item RepoBench. (2023). \emph{RepoBench: Benchmarking repository-level code auto-completion systems}. arXiv. \url{https://arxiv.org/abs/2306.03091}
\item METR. (2025, July 10). \emph{Measuring the impact of early-2025 AI on experienced open-source developer productivity}. \url{https://metr.org/blog/2025-07-10-early-2025-ai-experienced-os-dev-study/}
\item OpenAI. (2024, October 3). \emph{Introducing canvas}. \url{https://openai.com/index/introducing-canvas/}
\item CIO. (2025, September 23). Demand for junior developers softens as AI takes over. \url{https://www.cio.com/article/4062024/demand-for-junior-developers-softens-as-ai-takes-over.html}
\item Hampole, M., Papanikolaou, D., Schmidt, L. D. W., \& Seegmiller, B. (2025). \emph{Artificial intelligence and the labor market} (NBER Working Paper 33509). \url{https://www.nber.org/papers/w33509}
\item Hyman, B., Lahey, B., Ni, K., \& Pilossoph, L. (2025). \emph{How retrainable are AI-exposed workers?} Federal Reserve Bank of New York Staff Reports, No. 1165. \url{https://www.newyorkfed.org/research/staff\_reports/sr1165}
\item Selwyn, N. (2025). \emph{The ethics of AI or techno-solutionism?}. \emph{Journal of Education Policy}. \url{https://www.tandfonline.com/doi/abs/10.1080/01425692.2025.2502808}
\item Thomson Reuters. (2024, March 26). \emph{AI washing and SEC enforcement}. \url{https://www.thomsonreuters.com/en-us/posts/investigation-fraud-and-risk/ai-washing-enforcement/}
\item PwC. (2024). \emph{Global investor survey 2024}. \url{https://www.pwc.com/th/en/press-room/press-release/2024/press-release-26-12-24-en.html}
\item OECD. (2025). \emph{Competition in artificial intelligence infrastructure}. \url{https://www.oecd.org/en/publications/competition-in-artificial-intelligence-infrastructure\_623d1874-en.html}
\item Ishkhanyan, A. (2025). The sovereignty-internationalism paradox in AI governance: Digital federalism and global algorithmic control. \emph{Discover Artificial Intelligence, 5}, Article 123. \url{https://link.springer.com/article/10.1007/s44163-025-00374-x}
\item Papyshev, G., \& Chan, K. J. D. (2026). AI regulatory strategies for digital sovereignty: The role of geopolitics and technological disparities. \emph{Electronic Markets, 36}, Article 8. \url{https://doi.org/10.1007/s12525-025-00870-z}
\item CNBC. (2025a, May 13). \emph{Microsoft laying off about 6,000 people, or 3\% of its workforce}. \url{https://www.cnbc.com/2025/05/13/microsoft-is-cutting-3percent-of-workers-across-the-software-company.html}
\item Amazon. (2025, June 17). \emph{Message from CEO Andy Jassy: Some thoughts on Generative AI}. About Amazon. \url{https://www.aboutamazon.com/news/company-news/amazon-ceo-andy-jassy-on-generative-ai/}
\item CNBC. (2025c, June 26). \emph{AI is doing up to 50\% of the work at Salesforce, CEO Marc Benioff says}. \url{https://www.cnbc.com/2025/06/26/ai-salesforce-benioff.html}
\item Fortune. (2026, January 29). \emph{Top engineers at Anthropic, OpenAI say AI now writes 100\% of their code, with big implications for the future of software development jobs}. \url{https://fortune.com/2026/01/29/100-percent-of-code-at-anthropic-and-openai-is-now-ai-written-boris-cherny-roon/}
\item U.S. Senate Commerce Committee. (2026, March 3). \emph{Less Hype, More Help: AI That Improves Safety, Productivity, and Care}. \url{https://www.commerce.senate.gov/2026/3/less-hype-more-help-ai-that-improves-safety-productivity-and-care}
\item Siemens. (2026, March 3). \emph{Testimony before the U.S. Senate Commerce Subcommittee hearing “Less Hype, More Help: AI That Improves Safety, Productivity, and Care”}. \url{https://www.commerce.senate.gov/services/files/C9ACEB55-01EB-4F71-8E06-E6015C4C3886}
\item BLS. (2026, March 6). \emph{The Employment Situation - February 2026}. U.S. Bureau of Labor Statistics. \url{https://www.bls.gov/news.release/archives/empsit\_03062026.htm}
\item Terry, S. J. (2023). \emph{The macro impact of short-termism}. \emph{Econometrica, 91}(5), 1881-1912. \url{https://doi.org/10.3982/ECTA15420}
\item Green, F., Felstead, A., Gallie, D., Inanc, H., \& Jewson, N. (2016). The declining volume of workers' training in Britain. \emph{British Journal of Industrial Relations, 54}(2), 422-448. \url{https://doi.org/10.1111/bjir.12130}
\item Raisch, S., \& Krakowski, S. (2021). Artificial intelligence and management: The automation-augmentation paradox. \emph{Academy of Management Review, 46}(1), 192-210. \url{https://doi.org/10.5465/amr.2018.0072}
\item Manning, S. J., \& Aguirre, T. (2026). \emph{How adaptable are American workers to AI-induced job displacement?} (NBER Working Paper 34705). \url{https://www.nber.org/papers/w34705}
\item Brookings. (2025, May 16). \emph{AI labor displacement and the limits of worker retraining}. \url{https://www.brookings.edu/articles/ai-labor-displacement-and-the-limits-of-worker-retraining/}
\item Liu, N. F., Lin, K., Hewitt, J., Paranjape, A., Bevilacqua, M., Petroni, F., \& Liang, P. (2024). \emph{Lost in the middle: How language models use long contexts}. \emph{Transactions of the Association for Computational Linguistics, 12}, 157-173. \url{https://direct.mit.edu/tacl/article/doi/10.1162/tacl\_a\_00638/119630/Lost-in-the-Middle-How-Language-Models-Use-Long}
\item Google DORA. (2025). \emph{State of AI-assisted software development}. \url{https://dora.dev/dora-report-2025}
\item Sonar. (2026a, January 8). Sonar data reveals critical verification gap in AI coding. \url{https://www.sonarsource.com/company/press-releases/sonar-data-reveals-critical-verification-gap-in-ai-coding/}
\item GitHub. (2024, November 18). \emph{Does GitHub Copilot improve code quality? Here's what the data says}. The GitHub Blog. \url{https://github.blog/news-insights/research/does-github-copilot-improve-code-quality-heres-what-the-data-says/}
\item GAO. (2025). \emph{Information technology: Agencies need to plan for modernizing critical decades-old legacy systems} (GAO-25-107795). \url{https://files.gao.gov/reports/GAO-25-107795/index.html}
\item Villalobos, P., Ho, A., Sevilla, J., Besiroglu, T., Heim, L., \& Hobbhahn, M. (2024). \emph{Will we run out of data? Limits of LLM scaling based on human-generated data}. \emph{Epoch AI}. \url{https://epoch.ai/blog/will-we-run-out-of-data-limits-of-llm-scaling-based-on-human-generated-data}
\item Shumailov, I., Shumaylov, Z., Zhao, Y., Gal, Y., Papernot, N., \& Anderson, R. (2024). \emph{AI models collapse when trained on recursively generated data}. \emph{Nature, 631}, 755-759. \url{https://www.nature.com/articles/s41586-024-07566-y}
\item Kang, F., Ardalani, N., Kuchnik, M., Emad, Y., Elhoushi, M., Sengupta, S., Li, S.-W., Raghavendra, R., Jia, R., \& Wu, C.-J. (2025). \emph{Demystifying synthetic data in LLM pre-training: A systematic study of scaling laws, benefits, and pitfalls}. In \emph{Proceedings of the 2025 Conference on Empirical Methods in Natural Language Processing} (pp. 10739-10758). Association for Computational Linguistics. \url{https://aclanthology.org/2025.emnlp-main.544/}
\item Wan, A., Klyman, K., Kapoor, S., Maslej, N., Longpre, S., Xiong, B., Liang, P., \& Bommasani, R. (2025). \emph{The 2025 Foundation Model Transparency Index}. Stanford Center for Research on Foundation Models. \url{https://crfm.stanford.edu/fmti/December-2025/paper.pdf}
\item Reuters. (2025, January 31). Taiwan bans government agencies from using DeepSeek, citing security concerns. \emph{Taipei Times}. \url{https://www.taipeitimes.com/News/taiwan/archives/2025/01/31/2003831128}
\item CNBC. (2025, May 16). \emph{Musk's xAI says Grok's 'white genocide' posts resulted from change that violated 'core values'}. \url{https://www.cnbc.com/2025/05/15/musks-xai-grok-white-genocide-posts-violated-core-values.html}
\item Microsoft. (2025, January 21). \emph{Microsoft and OpenAI evolve partnership to drive the next phase of AI}. The Official Microsoft Blog. \url{https://blogs.microsoft.com/blog/2025/01/21/microsoft-and-openai-evolve-partnership-to-drive-the-next-phase-of-ai/}
\item Reuters. (2025, September 22). \emph{Nvidia to invest up to \$100 billion in OpenAI, linking two artificial intelligence titans}. Investing.com. \url{https://www.investing.com/news/stock-market-news/nvidia-to-invest-100-billion-in-openai-4249616}
\end{enumerate}
\endgroup
\normalsize
\end{document}